\documentclass[twocolumn,showpacs,aps,prb,superscriptaddress]{revtex4}

\usepackage{graphicx}
\usepackage{amsmath}
\usepackage{latexsym}

\begin{document}

\title{Coulomb Drag near the Metal-Insulator Transition in Two-Dimensions}

\author{R. Pillarisetty}
\affiliation{Department of Electrical Engineering, Princeton
University, Princeton, New Jersey 08544}
\author{Hwayong Noh}
\affiliation{Department of Electrical Engineering, Princeton
University, Princeton, New Jersey 08544}
\affiliation{Department
of Physics, Sejong University, Seoul 143-747, Korea}
\author{E. Tutuc}
\affiliation{Department of Electrical Engineering, Princeton
University, Princeton, New Jersey 08544}
\author{E.P. De Poortere}
\affiliation{Department of Electrical Engineering, Princeton
University, Princeton, New Jersey 08544}
\author{K. Lai}
\affiliation{Department of Electrical Engineering, Princeton
University, Princeton, New Jersey 08544}
\author{D.C. Tsui}
\affiliation{Department of Electrical Engineering, Princeton
University, Princeton, New Jersey 08544}
\author{M. Shayegan}
\affiliation{Department of Electrical Engineering, Princeton
University, Princeton, New Jersey 08544}

\date{\today}

\begin{abstract}
We studied the drag resistivity between dilute two-dimensional
hole systems, near the apparent metal-insulator transition. We
find the deviations from the $T^{2}$ dependence of the drag to be
independent of layer spacing and correlated with the metalliclike
behavior in the single layer resistivity, suggesting they both
arise from the same origin. In addition, layer spacing dependence
measurements suggest that while the screening properties of the
system remain relatively independent of temperature, they weaken
significantly as the carrier density is reduced. Finally, we
demonstrate that the drag itself significantly enhances the
metallic $T$ dependence in the single layer resistivity.
\end{abstract}
\maketitle

\section{Introduction}

Two-dimensional (2D) electron transport in semiconductor
heterostructures has provided a rich venue for the study of
electron interaction physics, exhibiting such exotic states as the
fractional quantum Hall liquid\cite{fqhe}. Here strong
interactions, induced by a large perpendicular magnetic field,
stabilize this non-Fermi liquid state. Recently, much attention in
the field has focused upon dilute 2D systems, which are
characterized by large ratios of carrier interaction energy to
kinetic energy ($r_{s} > 10$). These systems exhibit an anomalous
metalliclike behavior and an apparent metal-insulator
transition\cite{mit}, contradictory to the scaling theory of
localization\cite{scaling}. To date, the origin of the metallic
behavior is unclear, with several fundamental questions regarding
this regime still unanswered. Among the most important of these
are the nature of the many-body correlations and screening
properties in this dilute limit. To gain insight into both of
these issues, we have performed frictional drag measurements.

Drag measurement\cite{TJ} allows one to directly study
carrier-carrier interactions. These experiments are performed, on
double layer systems, by driving a current ($I_{D}$) in one layer,
and measuring the potential ($V_{D}$), which arises in the other
layer due to momentum transfer. The drag resistivity ($\rho_{D}$),
given by $V_{D}/I_{D}$, is directly proportional to the interlayer
carrier-carrier scattering rate. Furthermore, the layer spacing
($d$) dependence of the drag provides a powerful probe of both the
interlayer correlations and the screening properties, which exist
in the system. Any change in either of these properties will
manifest itself in the layer spacing dependence of the drag.
Recently, the drag was measured between low density hole
systems\cite{ravi}, with $r_{s}$ approximately ranging from 10 to
20\cite{note1}. In this regime, $\rho_{D}$ showed a 2 to 3 orders
of magnitude enhancement over the theory for weakly interacting
systems\cite{boltzmann}, and the corresponding low density
electron results\cite{2kf}. In addition, an anomalous temperature
($T$) dependence, which could not be explained in light of
previously studied drag processes, was observed, with $\rho_{D}$
exhibiting a greater than $T^{2}$ dependence at low temperatures.
Upon further increase of $T$, a crossover to a weaker than $T^{2}$
dependence was observed. It was proposed that these deviations
arose from many-body correlations in such a strongly interacting
regime\cite{ravi,hwang}.

In this article, we study the density, temperature, and layer
separation dependence of the drag between strongly interacting 2D
hole systems, in the vicinity of the apparent metal-insulator
transition. Using this data, we show that the deviations from the
$T^{2}$ dependence of the drag are not a result of a phonon
mediated drag process\cite{TJ2,phonon}, but rather arise from a
novel mechanism related to the large $r_{s}$ value of the system.
We find these deviations to be independent of layer spacing,
implying that they arise from an intralayer correlation effect.
Furthermore, we find that the deviations to the $T^{2}$ dependence
of the drag are correlated with the metalliclike $T$ dependence in
the single layer resistivity, suggesting that both anomalies have
the same origin. Our layer spacing dependence data imply that
while the screening properties in this regime remain relatively
independent of temperature, they weaken significantly as the
carrier density is reduced towards the metal-insulator transition.
Finally, we demonstrate that the Coulomb drag effect itself can
significantly enhance the metallic $T$ dependence in the single
layer resistivity.

\section{Experimental Details}

Four different samples were used in this study. Each sample
contains a double quantum well structure, consisting of two Si
doped p-type GaAs quantum wells separated by a pure AlAs
barrier\cite{note2}, which was grown by molecular beam epitaxy on
a (311)A GaAs substrate. Sample A, which was used in an earlier
study\cite{ravi}, has an average grown layer density and center to
center layer separation of $2.5\times10^{10}$ cm$^{-2}$ and 300
\AA, respectively. Samples B and C were similar to sample A, with
the exception of having different center to center layer
separations of 225 and 450 \AA, respectively. Sample D was a
higher density sample, having an average grown layer density and
center to center layer separation of $7.0\times10^{10}$ cm$^{-2}$
and 275 \AA, respectively. Explicit details of the parameters for
each of these four samples are listed in Table I. The samples were
processed allowing independent contact to each of the two layers,
using a selective depletion scheme\cite{ic}. In addition, both
layer densities are independently tunable using evaporated
metallic gates.

\begin{table}[b]
\caption{\label{tab1} Sample parameters. Mobilities quoted at 300
mK.}
\begin{ruledtabular}
\begin{tabular}{lcccccccc}
Sample                            & & A & B  & C & D\\
\hline
Average layer density [$\times10^{10}$ cm$^{-2}$]         & &2.5 &2.5 &2.5 &7.0\\
Top layer mobility [$\times10^{5}$ cm$^{2}$/Vs]         & &1.5 &1.9 &2.9 &5.7\\
Bottom layer mobility [$\times10^{5}$ cm$^{2}$/Vs]      & &1.5 &1.3 &1.7 &7.7\\
Quantum well thickness [\AA]                            & &150 &150 &150 &175\\
Barrier thickness [\AA]                                 & &150 &75 &300 &100\\
Center to center layer separation [\AA]                 & &300 &225 &450 &275\\
\end{tabular}
\end{ruledtabular}
\end{table}

The data presented in this paper were obtained using $^{3}$He and
dilution refrigerators. The densities in each layer were
determined by independently measuring Shubnikov-de Haas
oscillations. Drive currents between 50 pA to 10 nA were passed,
in the [$\bar{2}$33] direction, through one of the layers, while
the drag signal was measured in the other layer, using standard
lock-in techniques at 4 Hz. To ensure that no spurious sources
were contributing to our signal, all the standard consistency
checks associated with the drag technique were performed\cite{TJ}.

\begin{figure}[!t]
\begin{center}
\includegraphics[width=3.35in,trim=0.2in 0.2in 0.2in 0.2in]{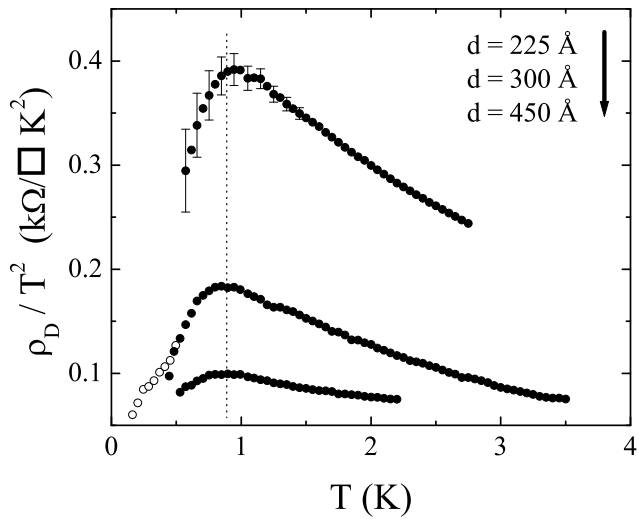}
\end{center}
\caption{\label{1}$\rho_{D}/T^{2}$ vs $T$ at
$p_{m}=2.5\times10^{10}$ cm$^{-2}$, for different $d$. For
clarity, data from the $d=$ 450 \AA\ sample has been multiplied by
a factor of 6.5. The dashed line marks the peak position. Open
circles measured in the dilution refrigerator.}
\end{figure}

\begin{figure}[!t]
\begin{center}
\includegraphics[width=3.35in,trim=0.2in 0.2in 0.2in 0.2in]{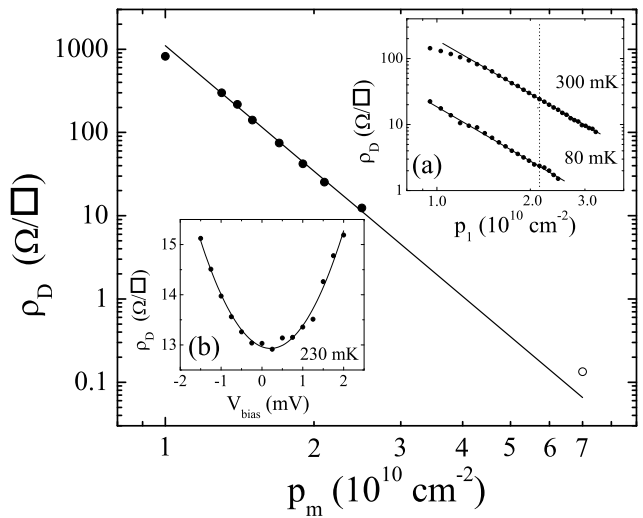}
\end{center}
\caption{\label{2}$\rho_{D}$ vs $p_{m}$ on log-log scale, at $T=$
300 mK. Data from Sample A ($d=$ 300 \AA) and D ($d=$ 275 \AA) are
shown by solid and open circles, respectively. The solid line is a
fit with slope -5. Inset (a): $\rho_{D}$ vs $p_{1}$ on log-log
scale, with $p_{2}=$ $2.15\times10^{10}$ cm$^{-2}$, at T = 80 and
300 mK. Solid lines are fits with slope close to -2.5. Dashed line
indicates matched density. Inset (b): $\rho_{D}$ vs $V_{bias}$.
$p_{m}=$ $2.1\times10^{10}$ cm$^{-2}$ and $T=$ 230 mK.}
\end{figure}

\section{Results and Discussion}

We begin our presentation of the data by first looking at the
temperature dependence of the drag in the dilute regime at
different layer spacings. This is presented in Fig.~\ref{1}, where
we plot $\rho_{D}/T^{2}$ vs $T$ at matched densities ($p_{m}$) of
$2.5\times10^{10}$ cm$^{-2}$, for $d=$ 225, 300, and 450 \AA. For
clarity, the data from the $d=$ 450 \AA\ sample has been
multiplied by a factor of 6.5. The error bar shown for the $d=$
225 \AA\ data is due to a small interlayer leakage contribution.
Note that both the $d=$ 225 and 450 \AA\ samples exhibit the same
qualitative behavior, showing a peak in $\rho_{D}/T^{2}$ vs $T$,
as was reported earlier\cite{ravi} at $d=$ 300 \AA. In addition,
the peak position in $\rho_{D}/T^{2}$ vs $T$, marked by the dashed
line, appears to be independent of layer separation. Although not
shown here, we also observed this at lower densities.

Before discussing this data, we would like to show that these
deviations from the $T^{2}$ dependence are not a result of a
phonon mediated drag process\cite{TJ2,phonon}. It has been well
established that a contribution from $2k_{F}$ phonon exchange to
the drag will produce deviations from the expected $T^{2}$
dependence, which are qualitatively similar to those shown in Fig
1. However, the fractional deviation from $T^{2}$ arising from a
phonon mediated process should decrease considerably as $d$ is
reduced. This stems from the fact that as $d$ is reduced, the
direct Coulomb component of the drag increases significantly,
whereas the phonon mediated component increases with a much weaker
$d$ dependence\cite{phonon}. The data in Fig 1. show that the
fractional deviation from $T^{2}$ is roughly independent of $d$.
Here, at all three layer spacings, $\rho_{D}/T^{2}$ changes
roughly 25 to 30 \%, from 0.5 to 2.0 K.

Furthermore, to conclusively rule out the phonon mediated process,
we present the relative density dependence of the drag at $d=$ 300
\AA, in Inset (a) of Fig.~\ref{2}. Here we have measured
$\rho_{D}$ as a function of one layer density ($p_{1}$), while
keeping the other layer density fixed at $p_{2}=
2.15\times10^{10}$ cm$^{-2}$. It has been demonstrated that the
phonon mediated component of the drag is strongly suppressed as
the layer densities are mismatched\cite{TJ2,phonon}. If $2k_{F}$
phonon exchange contributes to the drag in our samples, one would
expect to see some signature of a local maximum at matched density
in these density ratio measurements. However, as shown in the
inset, where $\rho_{D}$ is plotted on log-log scale against
$p_{1}$ at both $T=$ 80 and 300 mK, the drag exhibits no signature
of a local maximum at matched density, and is described by a
linear fit with slope close to -2.5, implying $\rho_{D} \propto
(p_{1}p_{2})^{-5/2}$ at low temperatures\cite{d450}. The deviation
from the fit at low density in the 300 mK trace is a consequence
of the density dependent crossover in the $T$ dependence of
$\rho_{D}$\cite{ravi}.

To provide yet further evidence against the presence of $2k_{f}$
phonon exchange in our samples, we have performed density
imbalance measurements\cite{2kf}. These measurements are performed
by measuring $\rho_{D}$ as a function of interlayer bias
($V_{bias}$). Interlayer bias transfers carriers between the
layers, while keeping the total density in both layers fixed. Such
a measurement provides a very sensitive probe as to the presence
of phonon mediated\cite{TJ2,phonon} (or direct\cite{2kf}) $2k_{F}$
scattering. A series expansion of the power law
$(p_{1}p_{2})^{-5/2}$ dependence deduced from our density ratio
measurements implies that $\rho_{D}$ will exhibit a quadratic
increase with interlayer bias. If $\rho_{D}$ contains any
significant contribution from $2k_{F}$ phonon exchange, this will
clearly lead to the drag decreasing as a function of $V_{bias}$.
In Inset (b) of Fig 2, we plot $\rho_{D}$ vs $V_{bias}$ at $T=$
230 mK. Here the layer densities are matched at $2.1\times10^{10}$
cm$^{-2}$ at zero bias. It is clear from the figure that
$\rho_{D}$ exhibits a quadratic increase with $V_{bias}$,
consistent with our density ratio results. Here $T/T_{F}=0.07$, so
thermal broadening of the Fermi surface is negligible. This data
conclusively shows that $2k_{F}$ phonon exchange\cite{TJ2,phonon}
(or direct $2k_{F}$ scattering processes\cite{2kf}) do not yield
any significant contribution to the drag at $d=$ 300 \AA.

In the main plot of Fig 2, we investigate the matched density
dependence of $\rho_{D}$. Here $\rho_{D}$ vs $p_{m}$ is plotted on
log-log scale for densities ranging from $p_{m}=$ 1.0 to
$2.5\times10^{10}$ cm$^{-2}$, taken from Sample A ($d=$ 300 \AA),
and at $p_{m}= 7.0\times10^{10}$, taken from Sample D ($d=$ 275
\AA). We find that this data is very well described by a linear
fit with slope of -5, implying that for this density range,
corresponding to $r_{s}=$ 5.1 to 13.5 (using m$^{*}=$
0.17m$_{e}$), $\rho_{D}$ follows a $p_{m}^{-5}$ dependence at low
temperature, consistent with our density ratio results. In the
Coulomb drag theory for weakly interacting
systems\cite{boltzmann}, which is successful in explaining the
results of drag experiments performed on high density electron
systems\cite{TJ} with $r_{s} \sim 1$, $\rho_{D}$ is expected to
scale as $p_{m}^{-3}$. The stronger density dependence we have
observed here is consistent with earlier reports\cite{ravi} that
the drag between low density holes exhibits a significant
enhancement over this simple theory, with the discrepancy
increasing as the density is reduced. In this context, what is of
particular interest is a recent experiment, on low density
electron double layer systems with $d=$ 280 \AA, which studied the
drag in a regime of intermediate interaction strength, with
$r_{s}$ ranging from about 2 to 4.3\cite{2kf}. This study also
found that the drag exhibited a power law dependence on the
matched electron density ($n$), and observed that $\rho_{D}
\propto n^{-4}$. This result suggests that there is a crossover in
the behavior of the drag from a $p_{m}^{-3}$ to a $p_{m}^{-5}$
dependence as the $r_{s}$ value of the system is increased. This
provides yet further evidence that the large enhancement of the
drag and the deviations from the $T^{2}$ dependence found in the
dilute regime arise from correlation effects.

It is important to note that the deviations we have observed could
arise from two different types of correlation effects: either
interlayer or intralayer correlations. The $T$ dependence data for
different $d$ shown in Fig 1 provides very useful information
regarding differentiating between these two effects. If the
deviations from the $T^{2}$ dependence arise from an interlayer
correlation effect, we should expect them to exhibit a significant
change as the layer spacing is varied. However, this is not what
is observed. Both the fractional deviation from $T^{2}$ and the
peak position are found to be independent of $d$, implying that
interlayer correlation effects are not playing a role here, and
that the deviations arise from something intrinsic within each of
the single layers themselves.

\begin{figure}[!t]
\begin{center}
\includegraphics[width=3.5in,trim=0.0in 0.0in 0.0in 0.35in]{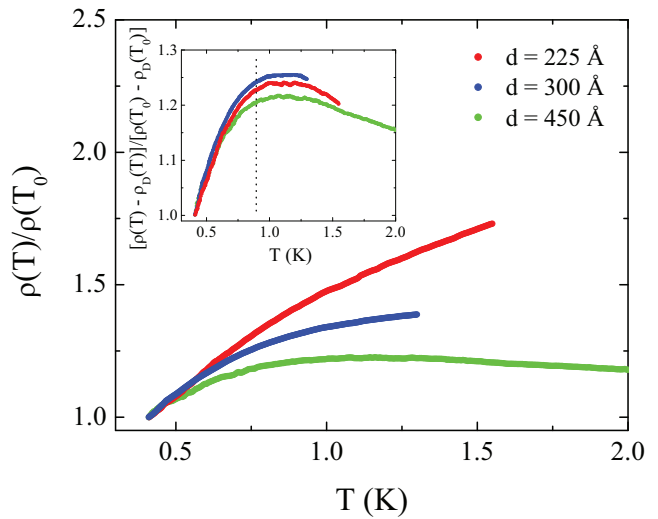}
\end{center}
\caption{\label{3}$\rho(T)/\rho(T_{0})$ vs $T$ at $p=$
2.5$\times10^{10} cm^{-2}$ for different $d$. Inset: [$\rho(T) -
\rho_{D}(T)]/[\rho(T_{0}) - \rho_{D}(T_{0})$] vs $T$. Dashed line
marks the peak position in $\rho_{D}/T^{2}$ vs $T$. $T_{0}=$ 400
mK.}
\end{figure}

At this point, it is appropriate to compare the deviations from
the $T^{2}$ dependence of the drag to the metalliclike behavior
observed in the individual layers of the samples used in this
study. Many believe that this strong metalliclike temperature
dependence arises from correlation effects, which exist in such a
large $r_{s}$ regime. By independently contacting one of the
layers in our double layer system, we can measure the temperature
dependence of the single layer resistivity. However, it is
important to realize that the resistivity obtained from such a
measurement is significantly enhanced due to the drag effect. Due
to its large magnitude, the drag effect will make a significant
contribution to the single layer resistivity. Since the drag has a
relatively strong temperature dependence, this will significantly
affect the strength of the metallic behavior obtained in a single
layer resistivity measurement. This is demonstrated in Fig. 3,
where we have plotted the normalized $T$ dependence of the single
layer resistivity, at $p=$ 2.5$\times10^{10}$ cm$^{-2}$, for
different $d$. For reference, the single layer resitivities at
$T=$ 400 mK are 1.48, 1.83, and 1.85 $k\Omega/\Box$ for the $d=$
225, 300, and 450 \AA\ samples, respectively. While all three of
the samples here have intrinsically similar properties
(resistivity and mobility), it is clear that the metallic behavior
becomes much stronger as $d$ is reduced, as should be expected
from the increasing drag contribution. Therefore, it is
appropriate to subtract the drag resistivity from the measured
single layer resistivity to obtain the intrinsic single layer
resistivity- that is the resistivity which would be measured if
the layers could be separated infinitely far apart. This is shown
in the inset, where we subtract $\rho_{D}$ from $\rho$ and plot
the normalized $T$ dependence of $\rho$ - $\rho_{D}$. Note here
that the metallic behavior in all three curves shows roughly the
same strength, confirming that the change in the metallic behavior
of the single layer resistivity, shown in the main plot, arises
from the drag contribution to the resistivity. The most striking
feature here is that the temperature at which the metallic
behavior is suppressed in $\rho$ - $\rho_{D}$ is at roughly the
same temperature at which the peak in $\rho_{D}/T^{2}$ vs $T$ is
observed, as shown by the dashed line in the inset. In addition,
in these three samples, the metallic temperature dependence
produces a roughly 20 to 25 \% change in $\rho$ - $\rho_{D}$. This
is extremely close to the 25 to 30 \% deviation from $T^{2}$,
which $\rho_{D}$ exhibits over the same temperature range. These
similarities seem to suggest that the deviations from the $T^{2}$
dependence of the drag are correlated with the anomalous metallic
behavior found in the single layer, and implies that they most
likely have the same origin.

The fact that both the single layer resistivity and the drag
exhibit similar anomalies is quite striking, since these are two
extremely different transport properties. Although there have been
numerous models attempting to explain the origin of the metallic
behavior, here we choose to analyze our data in light of one of
the most prominent: temperature dependent
screening\cite{dassarma}. In this model, the metallic behavior
arises from temperature dependent changes in the static screening
of ionized impurity potentials, which become important at large
$r_{s}$. As the temperature is increased from low $T$, the
screening weakens significantly, leading to the metalliclike
increase in the resistivity. It is possible to envision that such
a screening change could give rise to the enhancement to the
$T^{2}$ dependence of $\rho_{D}$ found at low $T$. However, here
we are concerned with the dynamic screening properties of the
system as it screens the interlayer Coulomb potential.

\begin{figure}[!t]
\begin{center}
\includegraphics[width=3.2in,trim=0.2in 0.2in 0.2in 0.2in]{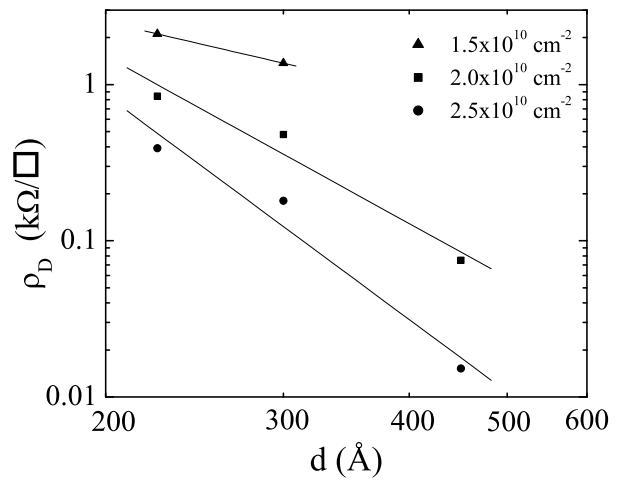}
\end{center}
\caption{\label{3}$\rho_{D}$ vs $d$, on log-log scale, for
$p_{m}=$ 1.5, 2.0, and $2.5\times10^{10}$ cm$^{-2}$ at T = 1.0 K.
Solid lines are the best linear fits of each data set.}
\end{figure}

\begin{figure}[!t]
\begin{center}
\includegraphics[width=3.5in,trim=0.0in 0.0in 0.0in 0.45in]{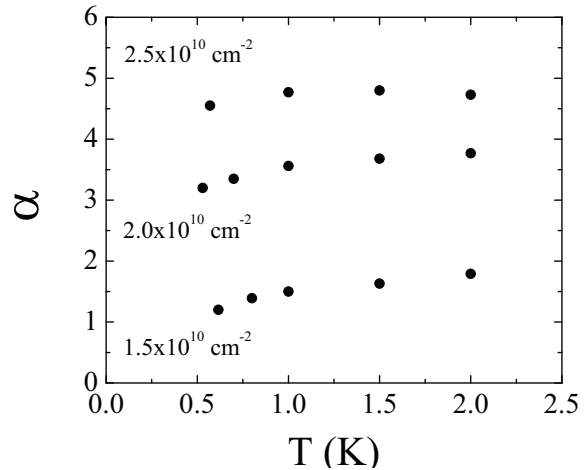}
\end{center}
\caption{\label{5}$\alpha$ vs $T$ for $p_{m}=$ 1.5, 2.0, and
$2.5\times10^{10}$ cm$^{-2}$. $\alpha$ deduced from fitting
$\rho_{D}$ vs $d$ to a $d^{-\alpha}$ fit.}
\end{figure}

If the deviations from the $T^{2}$ dependence of the drag arise
from a temperature dependent screening effect then this will
surely manifest itself in the layer spacing dependence of the
drag. The strength of the $d$ dependence of $\rho_{D}$ is
proportional to the screening strength of the interlayer Coulomb
potential. If the screening is strong, the interlayer potential
will drop rapidly with $d$ and $\rho_{D}$ will show a strong
decrease as $d$ is increased. On the other hand, if screening is
weak the interlayer potential will drop off much slower with $d$,
and $\rho_{D}$ will exhibit a much weaker decrease as $d$ is
increased. By investigating the strength of the $d$ dependence at
different temperatures, we can determine whether the dynamic
screening properties of the 2D system in this regime exhibit any
temperature dependent behavior. Using data from Samples A, B, and
C, we have studied $\rho_{D}$ as a function of $d$ at various
temperatures, for $p_{m}=$ 2.5, 2.0, and $1.5\times10^{10}$
cm$^{-2}$. A typical data set, obtained at $T=$ 1.0 K and plotted
on log-log scale, is shown in Fig 4. As expected, we observe that
$\rho_{D}$ shows a strong increase with decreasing layer spacing.
The solid lines are linear fits of each data set, and their slopes
correspond to the exponent, $\alpha$, where $\rho_{D} \propto
d^{-\alpha}$. Using additional data at different temperatures,
which are not shown, we can determine the strength of the $d$
dependence at each density, for different $T$. We present this
data in Fig 5, where we plot $\alpha$ as a function of temperature
for different $p_{m}$. We mention that the $\alpha$ found for
$p_{m}= 1.5\times10^{10}$ cm$^{-2}$ is deduced only using data at
$d=$ 225 and 300 \AA. Before discussing this data, we would first
like to mention that due to the narrow range of data we are not
making any quantitative claims on the values of these exponents.
On the other hand, a comparison of the relative change of $\alpha$
with temperature and density is perfectly valid and independent of
any of the systematic error in the $d$ values of our samples. The
first point we make is that at all three densities, the exponent
shows a weak increase with $T$, implying that the dynamic
screening properties in this regime are relatively independent of
temperature. If anything, the screening strengthens slightly as
$T$ is increased, the opposite of what should be expected from the
$T$ dependent screening model. Another point which is clear from
this data is that the strength of the $d$ dependence weakens
significantly as the carrier density is lowered. This implies that
the dynamic screening properties of the 2D system weaken
dramatically as the carrier density is reduced towards the
metal-insulator transition.

Finally, we would like to refer back to the single layer transport
data presented in Fig 3, which clearly shows that in the dilute
regime, the drag constitutes a significant fraction of the
resistivity. In contrast, at small $r_{s}$\cite{TJ} the drag
contribution to the resistivity is roughly 0.1 \% at these
temperatures. This demonstrates that in the dilute regime an
electron-electron (e-e) scattering process can yield a substantial
$T$ dependent correction to the resistivity, and suggests the
possibility that the metallic behavior seen in so many dilute 2D
systems might arise from an e-e scattering process. For example, a
theory on the interaction corrections\cite{zna} in this regime has
been shown to qualitatively explain the metalliclike temperature
dependence measured in experiments\cite{icexp}. In this theory the
disorder breaks the translational invariance of the system, and
allows e-e scattering to make a finite contribution to the
resistivity. Our data suggest that the magnitude of this
interaction correction could possibly constitute a significant
fraction of the resistivity. Another possibility is that the
electron system forms two channels, with only one carrying the
current, which has been suggested by local compressibility
measurements\cite{ilani}. Here interchannel e-e scattering
contributes to the resistance of the current carrying channel and
could possibly create the metallic $T$ dependence. This is
analogous to the double layer system, in which only one layer
carries current and the other layer causes dissipation via
interlayer e-e scattering.

\section{Conclusions}

In conclusion, we have shown the deviations from the $T^{2}$
dependence of the drag to be independent of layer spacing and
correlated with the metallic $T$ dependence in the single layer
resistivity, suggesting that both anomalies have the same origin.
Our studies of the strength of the layer spacing dependence of
$\rho_{D}$ imply that the dynamic screening properties in this
regime are relatively temperature independent. However, they
weaken significantly as the carrier density is reduced. Finally,
in this dilute regime, we have demonstrated that the drag effect
itself can significantly enhance the metallic behavior in the
single layer resistivity.

\section{Acknowledgments}

This research was funded by the NSF, a DURINT grant from the ONR,
and the DOE.


\begin{thebibliography}{90}
\bibitem{d450} For completness, we would like to mention that
density ratio measurements in the $d=$ 450 \AA\ sample exhibit
features near matched density, which indicate the presence of a
small phonon mediated drag contribution. However, we believe that
the deviation from the $T^{2}$ dependence in this sample arises
from the same mechanism that produces this deviation in the
samples with smaller $d$. This is since we find that the
deviations from the $T^{2}$ dependence at $d=$ 450 \AA\ are
unaffected by significantly mismatching the layer densities, along
with the fact that both the fractional deviation to $T^{2}$ and
the peak position in $\rho_{D}/T^{2}$ vs $T$ are the same as those
found at smaller layer spacing.
\end{thebibliography}
\end{document}